# Applying Algebraic Specifications on Digital Right Management Systems

Nikolaos Triantafyllou[1], Katerina Ksystra[1], Petros Stefaneas[2] and Panayiotis Frangos[1]

[1] School of Electrical and Computer Engineering, National Technical University of Athens,
Heroon Polytechniou 9, 15780 Zografou, Athens, Greece
{nitriant, katksy, pfrangos}@central.ntua.gr

[2] School of Applied Mathematical and Physical Sciences, National Technical University of Athens, Heroon Polytechniou 9, 15780 Zografou, Athens, Greece
petros@math.ntua.gr

**Abstract.** Digital Right Management (DRM) Systems have been created to meet the need for digital content protection and distribution. In this paper we present some of the directions of our ongoing research to apply algebraic specification techniques on mobile DRM systems.

## 1 Introduction

Digital Rights Management systems (DRMs) control many aspects of the life cycle of digital contents including consumption, management and distribution. Key component of such a system is the language in which the permissions on contents and constraints are expressed; these are called Right Expression Languages (RELs).In our paper we present some of our ongoing research directions aiming to address some of the DRM systems problems [1][2], with the help of algebraic specifications. So far our research has been focused on the DRM standard of the Open Mobile Alliance [3].

This paper is organized as follows: Section 2 gives the outline of an abstract syntax and its specification for OMA REL [4]. OMA, presents an algorithm that deals with multiple licenses referring to the same content. In section 3 this algorithm is formally specified in CafeOBJ algebraic specification language and a safety property is verified. This algorithm is not the optimal to use as it explained in [2]. In section 4 we present a redesign of this algorithm based on Order Sorted Algebra [5] and give a formal proof that this algorithm is correct using the methodology presented in [6]. Finally we give some first ideas towards what we believe can solve the interoperability problems of RELs, using the theory of Institutions [7].

## 2 Formal Semantics for OMA REL

We have given algebraic semantics to the OMA REL component dealing with expressing the permissions and constraints on the contents. To achieve this, we first created an abstract syntax for the language. Then we translated this syntax to the CafeOBJ specification language in order to use its rewriting as a tool for validation.

### 2.1 CafeOBJ in a nutshell

CafeOBJ [8] is an executable algebraic specification language, implementing equational logic by rewriting. Equations are treated as left to right rewrite rules. It can also be used as a powerful interactive theorem prover with the proof scores method [11]. With CafeOBJ each module defines a sort. A visible sort is a specification for an abstract data type. This is denoted with the name of the sort inside *[]*. Hidden sorts are used to specify state machines and this is denoted by enclosing the name of the sort in *[]*. Sort ordering is simply declared using <. Concerning hidden sorts there are two kinds of operators; *action operators,* which change the state of a machine, and *observation operators*, which observe a specific value in a particular state of the machine. Equations are denoted using the keyword *eq* and conditional equations using the keyword *ceq*. Finally modules can be imported to other modules by either protecting them or extending them.

### 2.2 Abstract Syntax and Specification in CafeOBJ

OMA REL [4] is an XML based language. The part of the language that is responsible for the expression of rights is called the agreement model. Inside this model the constraints and permission of the language are defined.

The abstract syntax we proposed, its specification and some case studies can be found in [9]. Here we will only demonstrate one example. Assume that Alice has purchased the following license: *Display content named contentID1 as many times as you like, and Display or Print the content named contentID2 as many times as you like.* The translation of the license in our abstract syntax is shown in figure 1.

Having specified the above abstract syntax as rewriting rules in CafeOBJ we can validate sets of licenses. The first step is to specify in a script the license of interest. Let us suppose that we want to declare the permissions allowed by the above license. In our specification this is done by declaring the permission set as:

```
eq ps1=add(True==>contentID2 print,add(True ==> content
ID2 display,add(True==>contentID1 display,em-permset))).
```

$agr := $ agreement

  about {ContentID1 ,ContentID2}

  with True $\xrightarrow{\quad}$ or[P1 ; P2 ; P3]

*where*

$P1 := $ True $\xrightarrow{\text{ContentID1}}$ *display*

$P2 := $ True $\xrightarrow{\text{ContentID2}}$ *display*

$P3 := $ True $\xrightarrow{\text{ContentID2}}$ *print*

```
<o-ex:asset o-ex:id="Asset-1">
  <o-ex:context>
    <o-dd:uid>ContentID1</o-dd:uid>
  </o-ex:context>
</o-ex:asset>
<o-ex:asset o-ex:id="Asset-2">
  <o-ex:context>
    <o-dd:uid>ContentID2</o-dd:uid>
  </o-ex:context>
<o-ex:permission>
  <o-ex:asset o-ex:idref="Asset-1"/>
  <o-ex:asset o-ex:idref="Asset-2"/>
  <o-dd:display/>
</o-ex:permission>
<o-ex:permission>
  <o-ex:asset o-ex:idref="Asset-2"/>
  <o-dd:print/>
</o-ex:permission>
```

**Fig. 1.** Abstract syntax and corresponding license of OMA REL

Since we have written the license in our model we can attain the permissions allowed by this license, through the `PermissionSET` operator, by the following equation `eq permissionSET = add( F(agr1,aboutset,emreset)`. Where F is an operator that returns a set of permissions from the `PermissionSet` whose constraints are met. After the permission set is created, it is easy to perform e-validation by simply asking the CafeOBJ compiler if the desired permission belongs to the permissions allowed by this license, using the following reduction `red Permitted(print,ebook,contentID2) in permissionSET`.

## 3 Verifying the OMA Rights Choice Algorithm

Together with the specification of OMA REL comes an algorithm that is responsible for choosing the most appropriate license to use, regarding a content, when there exist multiple licenses installed that refer to it. This algorithm has been formally specified in [10] and in addition, using the OTS/CafeOBJ [11] method the algorithm was proven to hold a safety property.

### 3.1 The OTS/CafeOBJ Method

An Observation Transition System (OTS) is a transition system that can be written in terms of equations. We assume there exists a universal state space, say Y. Formally, an OTS S is a triplet S = <O, I, T> where I is a subset of Y, the set of initial states of the machine and O is a set of observation operators. Each observer in O is an operator that takes a state of the system and possibly a series of other data type values (visible sorts) and returns a value of a data type that is characteristic to that state of the system.

| Safety Property for the OMA Rights Choice Algorithm |
|---|
| *Whenever a license is chosen for a content, then the licenses constraints are met at that specific time* |

**Table 1.** The safety property

Finally, T is the set of transition (or action) operators. Each transition takes as input a state of the system and again possibly a series of datatype values and returns a new state of the system. An OTS is transferred to CafeOBJ in a natural way. The state space corresponds to the values of a hidden sort. The initial states are denoted by a set of constants of the hidden sort. Observation operators are denoted as observers and transitions as action operators.

### 3.2 Formal Specification and Verification of the Algorithm

In [10] we have formally specified the Rights Choice Algorithm as an OTS written in CafeOBJ. The invariant property we have proved is can be seen in table 1. In order to prove such a property in CafeOBJ several steps need to be taken [11]. First you need to express the property as a predicate in CafeOBJ terms. Next, show that the predicate holds in any initial state. This is done by asking CafeOBJ to reduce the predicate term in an arbitrary initial state. Then we need to show that the property holds for any transition, the inductive step. Assuming that the predicate holds for an arbitrary state we ask CafeOBJ to reduce whether this implies that it holds for its successor state. The successor state is obtained by applying the transition rules to the above arbitrary state. CafeOBJ will either return true, false or an expression. If it returns true then the predicate holds on that step. When an expression is returned, this means that the machine cannot continue with the reductions. We must then assist CafeOBJ by case splitting the transition providing additional equations. If false is returned then we might need to find a lemma to discard this case or if this is not possible we are presented with a counter example. Following the OTS/CafeOBJ proof score the above property was verified in [10]. The proof of this property required the proof of five extra lemmas.

## 4 Proposing a New Algorithm and its Verification

There exist some cases where we end up losing execution rights by using the algorithm currently in use [2]. Indeed, let us consider the set of licenses seen on table 2. If the user decides to use his right "*listen to song A*", using the above algorithm the DRM agent will choose License 1. But by doing so, License 1 will become depleted since it contains the count constraint denoted by "once". This results in the user losing the right to ever listen to song B with this set of licenses. This would not occur if the agent had decided to use License 2 to execute the right to listen to song A.

This loss has been characterized by monotonicity of licenses in [2] and is proven that any algorithm attempting to solve this problem as is, will be NP-complete. Our approach is based on Order Sorted Algebra [5].

| Installed Licenses on a DRM agent | |
|---|---|
| *License1: "you may listen to songs A or B once before the end of the month".* | *License2: "you may listen to songs A or D ten times."* |

Table 2. A set of installed license that can cause a loss of rights

| Liveness Property for the OMA Rights Choice Algorithm |
|---|
| *If a right belongs to the installed licenses and is colored white leads to it being colored black.* |

Table 3. Liveness property describing the no loss of rights

We point out that licenses, as data types, can be represented by ordered sorts [12]. Next we identified that this loss of rights can only occur in some special cases. To capture this we inserted *Labels* on licenses that denote the following three things; Firstly, if the license contains one or more permissions, secondly the dominant constraint based on the original algorithm and finally if the license only allows one more execution. These labels allow us to provide an ordering on licenses that is used to determine what license to choose so that no loss will occur, while respecting the ordering on constraints in the original algorithm. The algorithm can be seen in detail in [12] together with several case studies conducted on a Java implementation of it.

### 4.1 Verification of the New Algorithm

We have proved that our new algorithm does not cause the same loss of rights as the algorithm currently in use. The proving procedure has been broken down into the following steps. First we created a specification of our algorithm as an OTS in CafeOBJ. Next we constructed an OTS, modeling the behavior of installed licenses on a DRM agent, meaning how they evolve when the user executes rights. The two OTSs where composed behaviorally as described in [13] yielding a new OTS. In order describe and to prove the desired property we added to the OTS a coloring on rights via an observer. Initially all rights are white (unused). A right is colored black (used) in two cases. Firstly, if the right corresponds to user request and the algorithm chooses the license containing this right as the optimal. Secondly, a right, say B, should be colored black if the user makes a request, say A different then B, but A only belongs to the license that contains B and that license becomes depleted after the execution of the request A.

The property describing the no loss of rights condition can be seen in table 3. This is a *Liveness* property and particularly a *leads-to* property [6]. The proof followed the methodology of [6]. The *lead-to* predicate was broken down into two *ensure* predicates of the form *p ensure q*, with p and q predicates. These types of properties require proving the "*unless case*; *p unless q*" and the "*eventually case*; *p eventually q*". For the first we need to prove that all of the transitions preserve the predicate; *(p(s) and $\neg$ q(s)) $\rightarrow$ (p(s') or q(s'))*. While for the second we need to show that there exists an instance of a transition where; *(p(s) and $\neg$ q(s)) $\rightarrow$ q(s')* holds. Where *s* a state of the OTS and *s'* is derived from *s* by applying a transition rule

## 6 Conclusions

We have presented some of our work on using algebraic specifications for mobile DRM systems. We have shown how various techniques from rewriting to theorem proving can help solving some of the problems on the field and also provide insights that can lead to the development of new technologies as with the proposed algorithm.

One of the main concerns with DRM is interoperability. There exist many different REL and DRM systems that do not work together , so at the moment it is usually not possible to transfer licenses from one environment (mobile)  to another (media player).We have started to address this problem by defining an Institution ([7]) for OMA REL. After defining Institutions for other commonly used RELs we intend to define a mechanism for translating licenses from one system to another via Institution morphisms in a way that preserves the meaning of the license, i.e. under a set of constraints a set of permissions is allowed, without it being a strict syntactic translation but rather a semantic one.